\documentclass[prl,twocolumn,superscriptaddress,showpacs,preprintnumbers,amsmath,amssymb]{revtex4}

\usepackage{graphicx}
\usepackage{epsfig}
\usepackage{dcolumn}
\usepackage{bm}

\newcommand{\lum}{{\cal L}}

\newcommand{\BR}{{\cal B}}

\newcommand{\jpsi}{J/\psi}
\newcommand{\EE}{e^+e^-}

\newcommand{\pp}{\pi^+\pi^-}

\newcommand{\kk}{K^+K^-}

\newcommand{\beq}{\begin{equation}}
\newcommand{\eeq}{\end{equation}}
\newcommand{\beqy}{\begin{eqnarray}}
\newcommand{\eeqy}{\end{eqnarray}}
\newcommand{\bitm}{\begin{itemize}}
\newcommand{\eitm}{\end{itemize}}

\begin{document}

\preprint{} \preprint{\vbox{
        \hbox{Belle Preprint 2012-14}
        \hbox{KEK   Preprint 2012-6}
         }}
\title{\quad\\[0.5cm]
First observation of exclusive $\Upsilon(1S)$ and $\Upsilon(2S)$
decays into light hadrons}




\affiliation{University of Bonn, Bonn}
\affiliation{Budker Institute of Nuclear Physics SB RAS and Novosibirsk State University, Novosibirsk 630090}
\affiliation{Faculty of Mathematics and Physics, Charles University, Prague}
\affiliation{University of Cincinnati, Cincinnati, Ohio 45221}
\affiliation{Department of Physics, Fu Jen Catholic University, Taipei}
\affiliation{Gifu University, Gifu}
\affiliation{Gyeongsang National University, Chinju}
\affiliation{Hanyang University, Seoul}
\affiliation{University of Hawaii, Honolulu, Hawaii 96822}
\affiliation{High Energy Accelerator Research Organization (KEK), Tsukuba}
\affiliation{Indian Institute of Technology Guwahati, Guwahati}
\affiliation{Indian Institute of Technology Madras, Madras}
\affiliation{Institute of High Energy Physics, Chinese Academy of Sciences, Beijing}
\affiliation{Institute of High Energy Physics, Vienna}
\affiliation{Institute of High Energy Physics, Protvino}
\affiliation{Institute for Theoretical and Experimental Physics, Moscow}
\affiliation{J. Stefan Institute, Ljubljana}
\affiliation{Kanagawa University, Yokohama}
\affiliation{Institut f\"ur Experimentelle Kernphysik, Karlsruher Institut f\"ur Technologie, Karlsruhe}
\affiliation{Korea Institute of Science and Technology Information, Daejeon}
\affiliation{Korea University, Seoul}
\affiliation{Kyungpook National University, Taegu}
\affiliation{\'Ecole Polytechnique F\'ed\'erale de Lausanne (EPFL), Lausanne}
\affiliation{Faculty of Mathematics and Physics, University of Ljubljana, Ljubljana}
\affiliation{Luther College, Decorah, Iowa 52101}
\affiliation{University of Maribor, Maribor}
\affiliation{Max-Planck-Institut f\"ur Physik, M\"unchen}
\affiliation{University of Melbourne, School of Physics, Victoria 3010}
\affiliation{Graduate School of Science, Nagoya University, Nagoya}
\affiliation{Kobayashi-Maskawa Institute, Nagoya University, Nagoya}
\affiliation{Nara Women's University, Nara}
\affiliation{National Central University, Chung-li}
\affiliation{National United University, Miao Li}
\affiliation{Department of Physics, National Taiwan University, Taipei}
\affiliation{H. Niewodniczanski Institute of Nuclear Physics, Krakow}
\affiliation{Nippon Dental University, Niigata}
\affiliation{Niigata University, Niigata}
\affiliation{Osaka City University, Osaka}
\affiliation{Pacific Northwest National Laboratory, Richland, Washington 99352}
\affiliation{Panjab University, Chandigarh}
\affiliation{Research Center for Electron Photon Science, Tohoku University, Sendai}
\affiliation{University of Science and Technology of China, Hefei}
\affiliation{Seoul National University, Seoul}
\affiliation{Sungkyunkwan University, Suwon}
\affiliation{School of Physics, University of Sydney, NSW 2006}
\affiliation{Tata Institute of Fundamental Research, Mumbai}
\affiliation{Excellence Cluster Universe, Technische Universit\"at M\"unchen, Garching}
\affiliation{Toho University, Funabashi}
\affiliation{Tohoku Gakuin University, Tagajo}
\affiliation{Tohoku University, Sendai}
\affiliation{Department of Physics, University of Tokyo, Tokyo}
\affiliation{Tokyo Institute of Technology, Tokyo}
\affiliation{Tokyo Metropolitan University, Tokyo}
\affiliation{Tokyo University of Agriculture and Technology, Tokyo}
\affiliation{CNP, Virginia Polytechnic Institute and State University, Blacksburg, Virginia 24061}
\affiliation{Wayne State University, Detroit, Michigan 48202}
\affiliation{Yamagata University, Yamagata}
\affiliation{Yonsei University, Seoul}
  \author{C.~P.~Shen}\affiliation{Graduate School of Science, Nagoya University, Nagoya} 
  \author{C.~Z.~Yuan}\affiliation{Institute of High Energy Physics, Chinese Academy of Sciences, Beijing} 
  \author{T.~Iijima}\affiliation{Kobayashi-Maskawa Institute, Nagoya University, Nagoya}\affiliation{Graduate School of Science, Nagoya University, Nagoya} 
  \author{I.~Adachi}\affiliation{High Energy Accelerator Research Organization (KEK), Tsukuba} 
  \author{H.~Aihara}\affiliation{Department of Physics, University of Tokyo, Tokyo} 
  \author{D.~M.~Asner}\affiliation{Pacific Northwest National Laboratory, Richland, Washington 99352} 
  \author{T.~Aushev}\affiliation{Institute for Theoretical and Experimental Physics, Moscow} 
  \author{A.~M.~Bakich}\affiliation{School of Physics, University of Sydney, NSW 2006} 
  \author{A.~Bay}\affiliation{\'Ecole Polytechnique F\'ed\'erale de Lausanne (EPFL), Lausanne} 
  \author{K.~Belous}\affiliation{Institute of High Energy Physics, Protvino} 
% ...
\collaboration{The Belle Collaboration}

\begin{abstract}

Using samples of 102 million $\Upsilon(1S)$ and 158 million
$\Upsilon(2S)$ events collected with the Belle detector, we study
exclusive hadronic decays of these two bottomonium resonances
to the three-body final states $\phi \kk$, $\omega \pp$ and
$K^{\ast 0}(892) K^- \pi^+ $, and to the two-body
Vector-Tensor states ($\phi f_2'(1525)$, $\omega f_2(1270)$, $\rho
a_2(1320)$ and $K^{\ast 0}(892) \bar{K}_2^{\ast 0}(1430) $) and
Axial-vector-Pseudoscalar ($K_1(1270)^+ K^-$, $K_1(1400)^+
K^- $ and $b_1(1235)^+ \pi^- $) states.
Signals are observed for the first time in the
$\Upsilon(1S) \to \phi K^+ K^-$, $\omega \pi^+ \pi^-$,
$K^{\ast 0} K^- \pi^+$, $K^{\ast0} K_2^{\ast 0}$ and
$\Upsilon(2S) \to \phi K^+ K^-$, $K^{\ast 0} K^- \pi^+$ decay modes.
Branching fractions are determined for all the processes, while 90\%
confidence level upper limits are established  on
the branching fractions for the modes with a statistical significance less
than $3\sigma$. The ratios of the branching
fractions of $\Upsilon(2S)$ and $\Upsilon(1S)$ decays into the same
final state are used to test a perturbative QCD prediction for OZI suppressed bottomonium decays.

\end{abstract}

\pacs{13.25.Gv, 14.40.Pq, 12.38.Qk}

\maketitle


Although around 80\% of the $\Upsilon(1S)$ decays and 60\% of the $\Upsilon(2S)$
decays are expected to result in hadronic final states via annihilation into
gluons~\cite{PDG, besson}, no single
exclusive mode has been reported~\cite{cleo-ydecay}. This situation is quite different
from the charmonium sector, where numerous channels have been measured
and used to test a variety of theoretical models. The OZI (Okubo-Zweig-Iizuka)~\cite{OZI} suppressed
decays of the $J/\psi$ and $\psi(2S)$ to hadrons proceed via the
annihilation of the quark-antiquark pair into three gluons or a
photon. For both cases, perturbative quantum chromodynamics
(pQCD) provides a relation for the ratios of branching fractions ($\cal B$)
for $\jpsi$ and $\psi(2S)$ decays~\cite{appelquist}
\begin{equation}
Q_{\psi}=\frac{{\cal B}_{\psi(2S) \to {\rm hadrons}}}{{\cal B}_{J/\psi \to {\rm hadrons}}}
=\frac{{\cal B}_{\psi(2S) \to e^+e^-}}{{\cal B}_{J/\psi \to
e^+e^-}} \approx 12\%,
\end{equation}
which is referred to as the ``12\% rule'' and is
expected to apply with reasonable accuracy to both inclusive
and exclusive decays. However, it is found to be severely violated for
$\rho\pi$ and other Vector-Pseudoscalar ($VP$) and
Vector-Tensor ($VT$) final states.
None of the many existing theoretical explanations that have been proposed have been
able to accommodate all the measurements reported to date~\cite{myw_review}.

A similar rule can be derived for OZI-suppressed bottomonium
decays, in which case we expect
 \begin{equation}
Q_{\Upsilon} =\frac{{\cal B}_{\Upsilon(2S) \to {\rm hadrons}}}{{\cal
B}_{\Upsilon(1S) \to {\rm hadrons}}} = \frac{{\cal B}_{\Upsilon(2S) \to
e^+e^-}}{{\cal B}_{\Upsilon(1S) \to e^+e^-}} = 0.77\pm 0.07.
 \end{equation}
This rule should hold better than the 12\% rule for charmonium decay,
since the bottomonium states have higher mass, pQCD and the
potential models should be more applicable, as has been demonstrated in
calculations of the $b\bar{b}$ meson spectrum.


In this Letter, we report studies of exclusive hadronic
decays of the $\Upsilon(1S)$ and $\Upsilon(2S)$ resonances to the
three-body final states $\phi \kk$, $\omega \pp$, and $K^{\ast
0}(892) K^- \pi^+$~\cite{charge} and two-body
$VT$ [$\phi f_2'(1525)\to \phi \kk$, $\omega f_2(1270) \to \omega \pp$, $\rho a_2(1320) \to \rho^0 \rho^+ \pi^-$, and
$K^{\ast 0}(892) \bar{K}_2^{\ast 0}(1430) \to K^{\ast 0}(892) K^- \pi^+$] and
Axial-vector-Pseudoscalar ($AP$) [$K_1(1270)^+ K^- \to \rho^0 K^+ K^-$, $K_1(1400)^+
K^-\to K^{\ast 0}(892) \pi^+ K^- $, and $b_1(1235)^+ \pi^- \to \omega \pi^+ \pi^-$] final states. This analysis is based on a
5.7~fb$^{-1}$  $\Upsilon(1S)$ data sample (102 million
$\Upsilon(1S)$ events), a 24.7~fb$^{-1}$ $\Upsilon(2S)$ data
sample (158 million $\Upsilon(2S)$ events), and a 89.4~fb$^{-1}$
continuum data sample collected at $\sqrt{s}=10.52$~GeV. Here,
$\sqrt{s}$ is the center-of-mass (C.M.) energy of the colliding
$e^+e^-$ system. The data are collected with the Belle
detector~\cite{Belle} operating at the KEKB asymmetric-energy
$\EE$ collider~\cite{KEKB}. The {\sc evtgen}~\cite{evtgen} generator is used to
simulate Monte Carlo (MC) events. For two-body decays, the angular
distributions are generated using the formulae in Ref.~\cite{DPNU}.
Inclusive $\Upsilon(1S)$ and $\Upsilon(2S)$ MC events, produced
using {\sc pythia}~\cite{pythia} with the same luminosity as real
data, are used to check for possible peaking backgrounds.



We require four reconstructed charged tracks with zero net charge.
For these tracks, the impact parameters perpendicular to and along
the beam direction with respect to the interaction point are
required to be less than 0.5~cm and 4~cm, respectively, and the
transverse momentum in the laboratory frame is restricted to be
higher than 0.1~$\hbox{GeV}/c$. For each charged track, we combine
information from different detector subsystems to form
a likelihood $\mathcal{L}_i$ for each particle species~\cite{pid}.

 A track with $\mathcal{R}_K =
\frac{\mathcal{L}_K} {\mathcal{L}_K + \mathcal{L}_\pi}> 0.6$ is
identified as a kaon, while a track with $\mathcal{R}_K<0.4$ is
treated as a pion. With this selection, the kaon (pion)
identification efficiency is about 89\% (92\%), while 6\% (9\%) of
kaons (pions) are misidentified as pions (kaons).
A similar likelihood ratio $\mathcal{R}_\mu$ is formed
for muon identification~\cite{MUID}. Except for the $\phi \kk$ final state
for which at least three charged tracks are required to be identified as kaons,
all other charged tracks are required to be positively
identified as pions or kaons. A small background
with muons is removed by requiring
$\mathcal{R}_\mu < 0.95$ for the pion candidates.

A neutral cluster in the electromagnetic calorimeter is reconstructed as a photon, if
it does not match the extrapolated position of any charged track and
its energy is greater than 40~MeV. A $\pi^0$ candidate is reconstructed
from a pair of photons. We perform a mass-constrained fit to the selected
$\pi^0$ candidate and require $\chi^2<6$.

\begin{figure}[htbp]
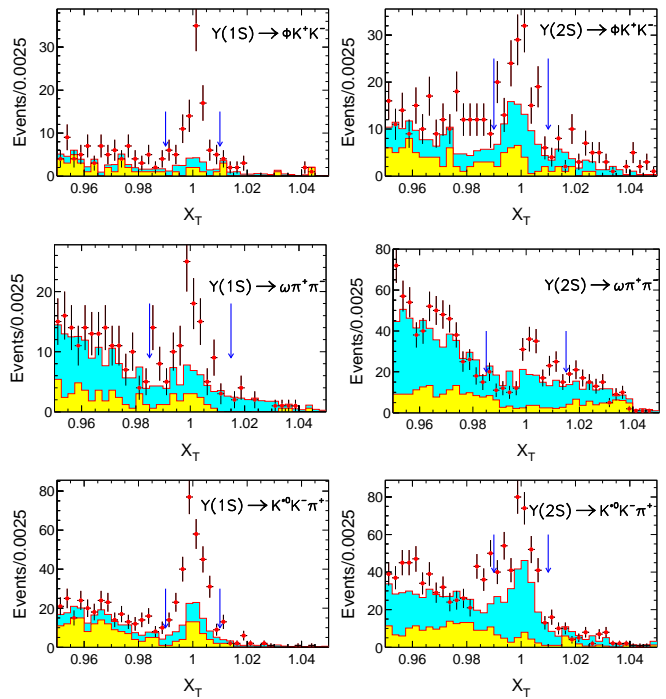

\includegraphics[height=2.9cm]{fig0a.epsi}\hspace{0.15cm}\vspace{0.1cm}
\includegraphics[height=2.9cm]{fig0b.epsi}\vspace{0.1cm}
\\
\includegraphics[height=2.9cm]{fig0c.epsi}\hspace{0.3cm}\vspace{0.1cm}
\includegraphics[height=2.9cm]{fig0d.epsi}\vspace{0.1cm}
\\
\includegraphics[height=2.9cm]{fig0e.epsi}\hspace{0.15cm}\vspace{0.1cm}
\includegraphics[height=2.9cm]{fig0f.epsi}\vspace{0.1cm}
\caption{\label{ecms}
Scaled total energy, $X_T$, distributions
from $\Upsilon(1S)$ and $\Upsilon(2S)$ decays to $\phi \kk$, $\omega \pp$
and $K^{\ast0}(892) K^- \pi^+$. The dots with error bars are from resonance data; the dark
shaded histograms are from normalized continuum contributions;
the light shaded  histograms are from inclusive $\Upsilon(1S)$ and $\Upsilon(2S)$ MC events
with signals removed. The arrows show the required signal region.}
\end{figure}

We impose an energy conservation requirement on $X_{T} =
\Sigma_h E_h/\sqrt{s}$, where $E_h$ is the energy of the
final state particle $h$ in the $\EE$ C.M. frame.
The ratio $X_T$ should lie in the range $0.985 \leq X_T \leq
1.015$ for channels with a $\pi^0$ in the final
state, and in the range $0.99 \leq X_T  \leq 1.01$ for other channels.
Figure~\ref{ecms} shows the $X_T$ distributions
from $\Upsilon(1S)$ and $\Upsilon(2S)$ decays to $\phi \kk$, $\omega \pp$
and $K^{\ast0}(892) K^- \pi^+$, together with expected backgrounds from
continuum processes and Y(1S) and Y(2S) decays.
Obvious signal candidates at $X_T\sim1$ can be seen.

\begin{figure}[htbp]
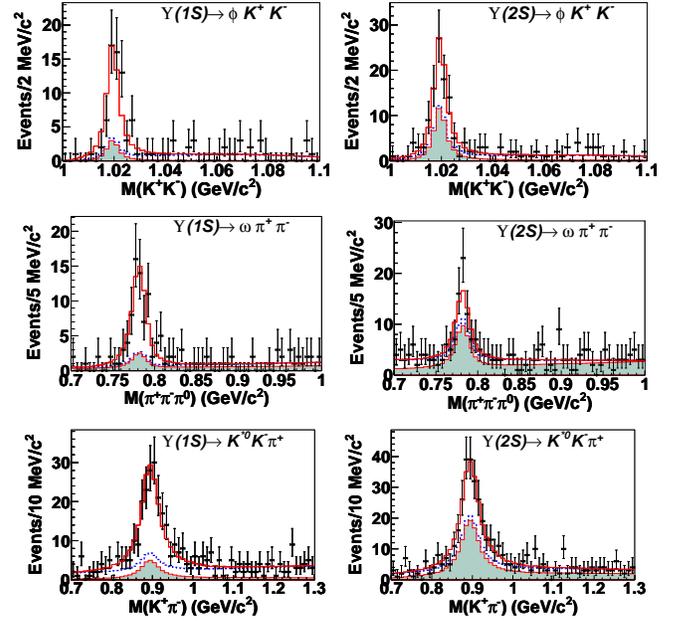

\includegraphics[height=4.2cm,angle=-90]{fig1a.epsi}\hspace{0.15cm}\vspace{0.1cm}
\includegraphics[height=4.2cm,angle=-90]{fig1b.epsi}\vspace{0.1cm}
\\
\includegraphics[height=4cm,angle=-90]{fig1c.epsi}\hspace{0.3cm}\vspace{0.1cm}
\includegraphics[height=4cm,angle=-90]{fig1d.epsi}\vspace{0.1cm}
\\
\includegraphics[height=4cm,angle=-90]{fig1e.epsi}\hspace{0.15cm}
\includegraphics[height=4cm,angle=-90]{fig1f.epsi}
\caption{\label{three-body-fit} The $\kk$ (top row), $\pp \pi^0$
(middle row), and $K^+ \pi^-$ (bottom row) invariant-mass
distributions for the final candidate events from
$\Upsilon(1S)$ (left column) and $\Upsilon(2S)$ (right column) three-body decays. Solid points
with error bars are data, open histograms show the best
fits, dashed curves are the total background estimates, and
shaded histograms are the normalized continuum background contributions.
}
\end{figure}

For three-body decay modes with a $\phi$ or $\omega$, $\phi$ ($\omega$)
candidates are selected with $\kk$ ($\pp \pi^0$) masses closest to
the nominal $\phi$ ($\omega$) mass~\cite{PDG}.  Figure~\ref{three-body-fit} shows the $\kk$, $\pp \pi^0$, and
$K^+ \pi^-$ invariant mass distributions for $\Upsilon(1S)$ and $\Upsilon(2S)$ to $\phi \kk$,
$\omega \pp$ and $K^{\ast 0}(892) K^- \pi^+$ candidates that survive the selection criteria
described above. Clear $\phi$,
$\omega$, and $K^{\ast 0}(892)$ signals are evident.

After the application of all of the
selection requirements, no peaking backgrounds from the $\Upsilon(1S)$ and $\Upsilon(2S)$
inclusive MC samples are found in the vector meson mass regions. Potential backgrounds due to particle misidentification,
from $\phi \pi^+ \pi^-$ for example, are estimated by selecting these events in the data
and normalizing them using measured misidentification probabilities.
Potential backgrounds from events with additional $\pi^0$'s are checked by
examining the recoil mass distribution from the measured final state. For $\omega \pi^+ \pi^-$, potential
background events from $\omega \eta'$ with
$\eta' \to \gamma \pi^+ \pi^-$ are explicitly reconstructed from data and estimated using $N\epsilon_1/\epsilon_2$,
where $N$ is the number of $\omega \eta'$ events in data and
$\epsilon_1$ and $\epsilon_2$ are the efficiencies after the $\omega \pi^+ \pi^-$ and
$\omega \eta'$ event selections, respectively. All of the above
backgrounds are found to be negligibly small. For $\omega \pi^+ \pi^-$, the
fraction of events with multiple combinations
is at the 3.5\% level due to multiple $\pi^0$  candidates; this is consistent with the MC
simulation and is taken into account in the efficiency determination.

The continuum background contribution is determined from the
data at $\sqrt{s}=10.52$ GeV and is extrapolated down to
the $\Upsilon(1S)$ and $\Upsilon(2S)$ resonances. For the
extrapolation, the scale factor,
$f_{{\rm scale}}$, is given by $\frac{\lum_{\Upsilon}}{\lum_{{\rm con}}}
\frac{\sigma_{\Upsilon}}{\sigma_{{\rm con}}}
\frac{\epsilon_{\Upsilon}}{\epsilon_{{\rm con}}}$, where
$\frac{\lum_{\Upsilon}}{\lum_{{\rm con}}}$,
$\frac{\sigma_{\Upsilon}}{\sigma_{{\rm con}}}$, and
$\frac{\epsilon_{\Upsilon}}{\epsilon_{{\rm con}}}$ are the ratios of
luminosity, cross sections, and efficiencies at the bottomonium masses and
continuum energy points. The $s$ dependence of the cross section is assumed to be $1/s$~\cite{foot}
and  the corresponding scale factor is 0.079 for the
$\Upsilon(1S)$ and 0.30 for the $\Upsilon(2S)$.

An unbinned simultaneous likelihood fit to the mass distributions is applied to extract the
signal and background yields in the $\Upsilon(1S)$ and continuum data samples and in the $\Upsilon(2S)$
and continuum data samples. The signal shapes are
obtained from MC simulations. In this fit,
a second-order Chebyshev polynomial background shape is used for the
$\Upsilon(1S)$/$\Upsilon(2S)$ decay backgrounds in addition to the normalized
continuum contribution.
The fit ranges and results
for the $K^+ K^-$, $\pp \pi^0$, and $K^+ \pi^-$ mass spectra are
shown in Fig.~\ref{three-body-fit} and Table~\ref{summary2}.

We determine a Bayesian 90\% confidence level (C.L.) upper limit on $N^{\rm sig}$
by finding the value $N^{\rm UP}_{\rm sig}$ such that
$
\frac{\int_{0}^{N^{\rm UP}_{\rm sig}} \mathcal{L} dN^{\rm sig}} {\int_{0}^{\infty} \mathcal{L} dN^{\rm sig}}=0.90,
$
where $N_{\rm sig}$ is the number of signal events and $\mathcal{L}$ is
the value of the likelihood as a function of $N_{\rm sig}$.
The statistical significance of the signal is estimated from the difference of the logarithmic likelihoods,
$-2\ln(\mathcal{L}_0/\mathcal{L}_{\rm max})$, taking into account the difference in the number of degrees of freedom
in the fits, where $\mathcal{L}_0$ and $\mathcal{L}_{\rm
max}$ are the likelihoods of the fits with and without signal, respectively.


\begin{table*}
\caption{Results for the $\Upsilon(1S)$ and $\Upsilon(2S)$ decays, where
$N^{\rm sig}$ is the number of fitted signal events, $N^{\rm
UP}_{\rm sig}$ is the upper limit on the number of signal events,
$\epsilon$ is the efficiency (\%),
$\Sigma$ is the statistical significance ($\sigma$), $\BR$ is the
branching fraction, $\BR^{\rm UP}$ is the upper limit on the
branching fraction,
$Q_{\Upsilon}$ is the
ratio of the $\Upsilon(2S)$ and $\Upsilon(1S)$ branching
fractions, and $Q_{\Upsilon}^{\rm UP}$ is the upper limit on the value of
$Q_{\Upsilon}$.
Branching fractions are in units of
$10^{-6}$ and upper limits are given at the 90\% C.L. The first error in $\BR$ and $Q_{\Upsilon}$ is statistical, and
the second systematic. } \label{summary2}
\begin{center}{
\scriptsize
\begin{tabular}{c|llcccl|llcccl|cc}
\hline
Channel & \multicolumn{6}{c|}{$\Upsilon$(1S)}  &  \multicolumn{6}{c|}{$\Upsilon$(2S)}   &   \\
        & $N^{\rm sig}$ & $N^{\rm UP}_{\rm sig}$& $\epsilon$ & $\Sigma$ & $\BR$
        &$\BR^{\rm UP}$  & $N^{\rm sig}$ & $N^{\rm UP}_{\rm sig}$ & $\epsilon$
        & $\Sigma$ &$\BR$ &$\BR^{\rm UP}$  &$Q_{\Upsilon}$ & $Q_{\Upsilon}^{\rm UP}$
\\
\hline \rule{0mm}{0.4cm}
 $\phi \kk$ & $56.3\pm 8.7$ &   & 47.9 &8.6 & $2.36\pm 0.37 \pm 0.29$  &
 & $58\pm12$ &   & 47.8 &6.5  & $1.58\pm0.33\pm 0.18$ &   & $0.67\pm0.18\pm 0.11$ &   \\
 $\omega \pp$  & $63.6\pm 9.5$ &   & 15.7 &8.5  & $4.46\pm0.67\pm0.72$  &
 &  $29 \pm 12 $ & 51  & 15.9 &2.5 & $1.32\pm0.54\pm0.45$  & 2.58   & $0.30\pm0.13 \pm 0.11$ & 0.55 \\
 $K^{\ast 0} K^- \pi^+$  & $173\pm 20$ &  & 28.7 &$11$ & $4.42\pm0.50 \pm 0.58$   &
 & $135 \pm 23 $ &  & 27.5 &6.4 & $2.32\pm0.40\pm 0.54$   &  & $0.52\pm0.11\pm 0.14$ &
\\\hline
 $\phi f_2'$  & $6.9\pm 3.9$ & 15  & 48.8 &2.1 &$0.64\pm 0.37 \pm 0.14$ & 1.63
 & $8.3\pm 6.0$ & 18  & 49.0 &1.6 &$0.50\pm0.36 \pm 0.19$ & 1.33 & $0.77\pm0.70 \pm 0.33$ & 2.54 \\
 $\omega f_2$  & $5.2\pm 4.0$ & 13  & 17.7 &1.5 &$0.57\pm0.44\pm 0.13$ & 1.79
 &  $-0.4\pm3.3$ & 6.1  &  17.5 &  &$-0.03\pm0.24\pm 0.01$ & 0.57  & $-0.06\pm0.42 \pm 0.02$ & 1.22  \\
 $\rho a_2$  & $29 \pm 11 $ & 49 & 17.4 & 2.7 &$1.15\pm0.47\pm 0.18$ & 2.24
 &  $ 10\pm 11$ & 30 & 17.3 & 0.9 & $0.27\pm0.28 \pm 0.14$ & 0.88  & $0.23\pm0.26 \pm 0.12$ & 0.82 \\
 $K^{\ast 0} \bar{K}_2^{\ast 0}$& $42.2\pm9.5$ &   & 30.8 &5.4 &$3.02\pm0.68\pm0.34$ &
 & $32\pm 11$ &   & 29.6 &3.3 &$1.53\pm0.52\pm0.19$ &  & $0.50\pm 0.21\pm 0.07$ &
\\\hline
 $K_1(1270)^+ K^-$ & $3.7\pm4.9$ & 13 & 23.6  & 0.8 &$0.54\pm0.72\pm 0.21$ & 2.41
 &  $11.0\pm4.4$ & 26 & 23.5 & 1.2 &$1.06\pm0.42\pm 0.32$ & 3.22  & $1.96\pm2.71 \pm 0.84$ & 4.73 \\
 $K_1(1400)^+ K^-$ & $23.8\pm8.2$ &  & 27.3 &3.3 &$1.02\pm0.35 \pm 0.22$ &
 & $9.2\pm8.2$ & 24  & 26.9 &0.5 &$0.26\pm0.23 \pm 0.09$ & 0.83 & $0.26\pm0.25\pm0.10$  & 0.77 \\
 $b_1(1235)^+\pi^-$  & $14.4\pm6.9$ & 28 & 16.7  & 2.4 &$0.47\pm0.22 \pm 0.13$ & 1.25
 & $1.2\pm3.5$ & 13 & 17.0  & 0.2 &$0.02\pm0.07 \pm 0.01$ & 0.40 & $0.05\pm0.16 \pm 0.03$ & 0.35 \\
\hline
\end{tabular}
}
\end{center}
\end{table*}

\begin{figure}[htbp]
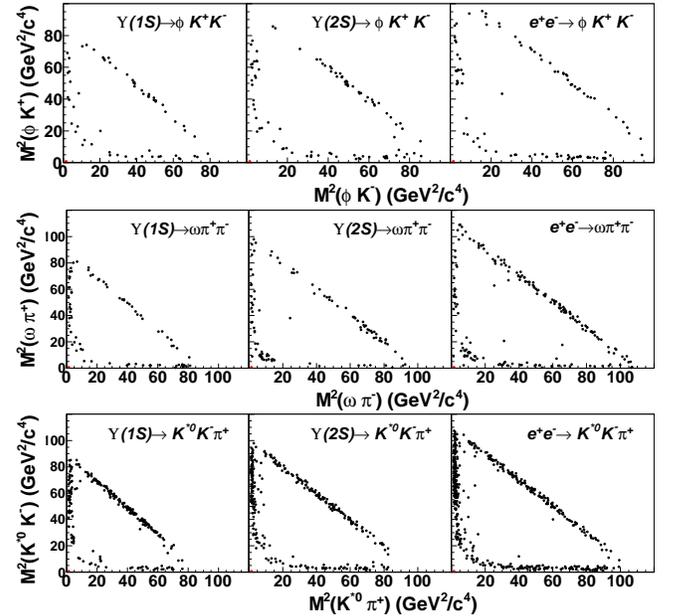

\includegraphics[height=8.5cm,angle=-90]{fig2a.epsi}\\
\includegraphics[height=8.5cm,angle=-90]{fig2b.epsi}\\
\includegraphics[height=8.5cm,angle=-90]{fig2c.epsi}
\caption{\label{dalitz} Dalitz plots of $\phi \kk$ (top row),
$\omega \pp$ (middle row) and $K^{\ast 0}(892) K^- \pi^+$ (bottom
row) three-body final states. Here, the left column is for $\Upsilon(1S)$ decays,
the middle column is for $\Upsilon(2S)$ decays, and the right column is for the continuum
 data. }
\end{figure}

After requiring $|M_{\kk}-m_{\phi}|<8$~MeV/$c^2$,
$|M_{\pp\pi^0}-m_{\omega}|<30$~MeV/$c^2$, and $|M_{K^+
\pi^-}-m_{K^{\ast 0}(892)}|<100$~MeV/$c^2$,
which contain around 95\% of the signal according to MC simulations,
the Dalitz plots for the
$\phi \kk$, $\omega \pp$, and $K^{\ast 0}(892) K^- \pi^+$ final
states are shown in Fig.~\ref{dalitz}, where $m_{\phi}$,
$m_{\omega}$, and $m_{K^{\ast 0}(892)}$ are the nominal
$\phi$, $\omega$, and $K^{\ast 0}(892)$ masses~\cite{PDG}.
Interestingly, the events accumulate near the phase space
boundary, reflecting the quasi-two-body nature of these decays.

To categorize the quasi-two-body decays into $VT$ or $AP$ final
states, we further require the angle between $V$ and $T$ ($A$ and
$P$) in the $\EE$ C.M. frame to be greater than 179 degrees for the
channels with a $\pi^0$, or 179.5 degrees for the other channels. The
combination with the minimum value of $\delta_{min} =
(M_1-m_{V})^2+(M_2-m_{T})^2$ is selected as the $V$ and $T$
candidate, where $M_1$ and $M_2$ are the invariant masses of the
$V$ and $T$ decay final-state particles, respectively. The same technique is used to
select the best $AP$ candidate.  This method
introduces negligible bias in the meson pair selection according to
MC simulation.

\begin{figure*}
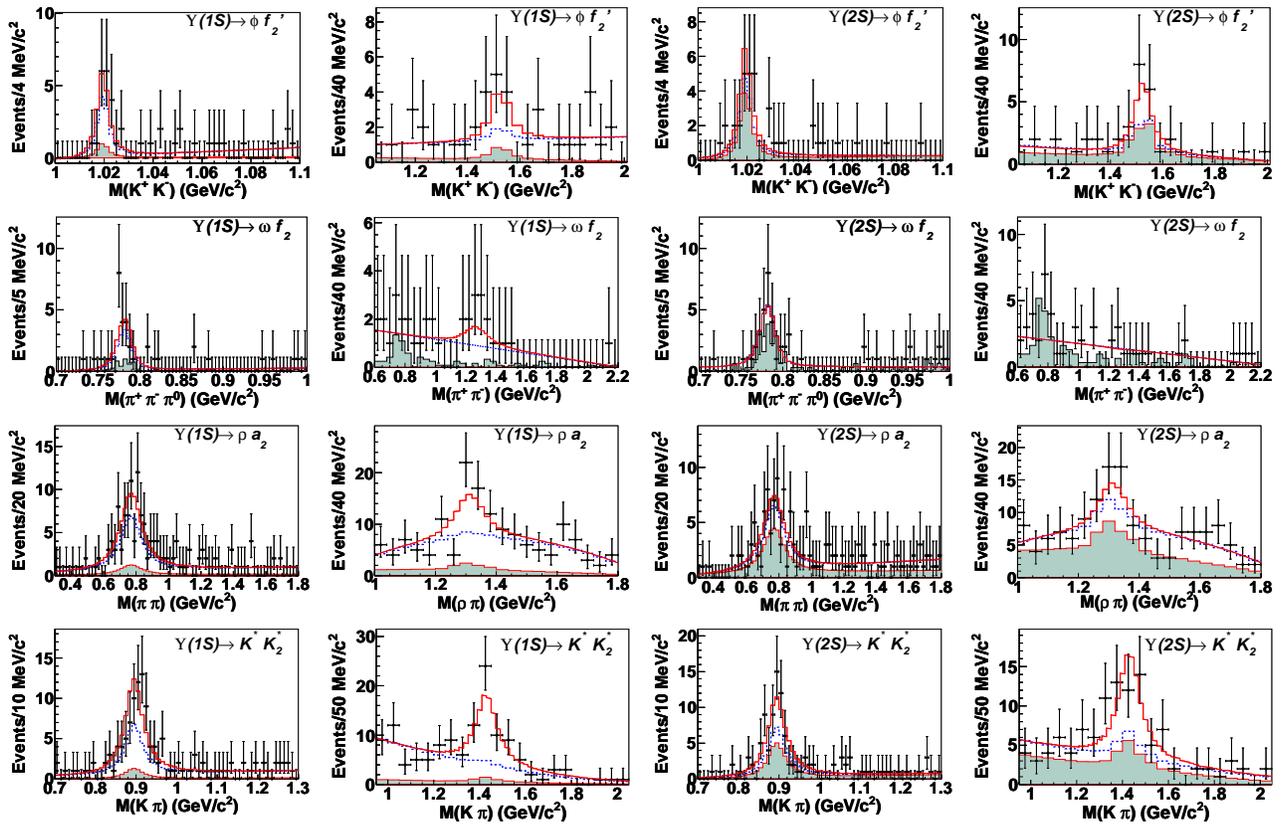

\includegraphics[height=4cm,angle=-90]{fig3a.epsi}\hspace{0.25cm}\vspace{0.05cm}
\includegraphics[height=4cm,angle=-90]{fig3b.epsi}\hspace{0.3cm}\vspace{0.05cm}
\includegraphics[height=4cm,angle=-90]{fig3c.epsi}\hspace{0.25cm}\vspace{0.05cm}
\includegraphics[height=4cm,angle=-90]{fig3d.epsi}\vspace{0.05cm}
\includegraphics[height=4cm,angle=-90]{fig3e.epsi}\hspace{0.25cm}\vspace{0.05cm}
\includegraphics[height=4cm,angle=-90]{fig3f.epsi}\hspace{0.3cm}\vspace{0.05cm}
\includegraphics[height=4cm,angle=-90]{fig3g.epsi}\hspace{0.25cm}\vspace{0.05cm}
\includegraphics[height=4cm,angle=-90]{fig3h.epsi}\vspace{0.05cm}
\includegraphics[height=4cm,angle=-90]{fig3i.epsi}\hspace{0.25cm}\vspace{0.05cm}
\includegraphics[height=4cm,angle=-90]{fig3j.epsi}\hspace{0.3cm}\vspace{0.05cm}
\includegraphics[height=4cm,angle=-90]{fig3k.epsi}\hspace{0.25cm}\vspace{0.05cm}
\includegraphics[height=4cm,angle=-90]{fig3l.epsi}\vspace{0.05cm}
\includegraphics[height=4cm,angle=-90]{fig3m.epsi}\hspace{0.25cm}\vspace{0.05cm}
\includegraphics[height=4cm,angle=-90]{fig3n.epsi}\hspace{0.3cm}\vspace{0.05cm}
\includegraphics[height=4cm,angle=-90]{fig3o.epsi}\hspace{0.25cm}\vspace{0.05cm}
\includegraphics[height=4cm,angle=-90]{fig3p.epsi}\vspace{0.05cm}
\caption{\label{vt-fit} The mass projections for
the vector and tensor meson candidates
from 2D fits to the events from $\Upsilon(1S)$ and $\Upsilon(2S)$
two-body decays ($VT$ modes). The open histograms show the results
of the 2D simultaneous fits, the dotted curves show the total
background estimates, and the shaded histograms are the
normalized continuum contributions.}
\end{figure*}

\begin{figure*}
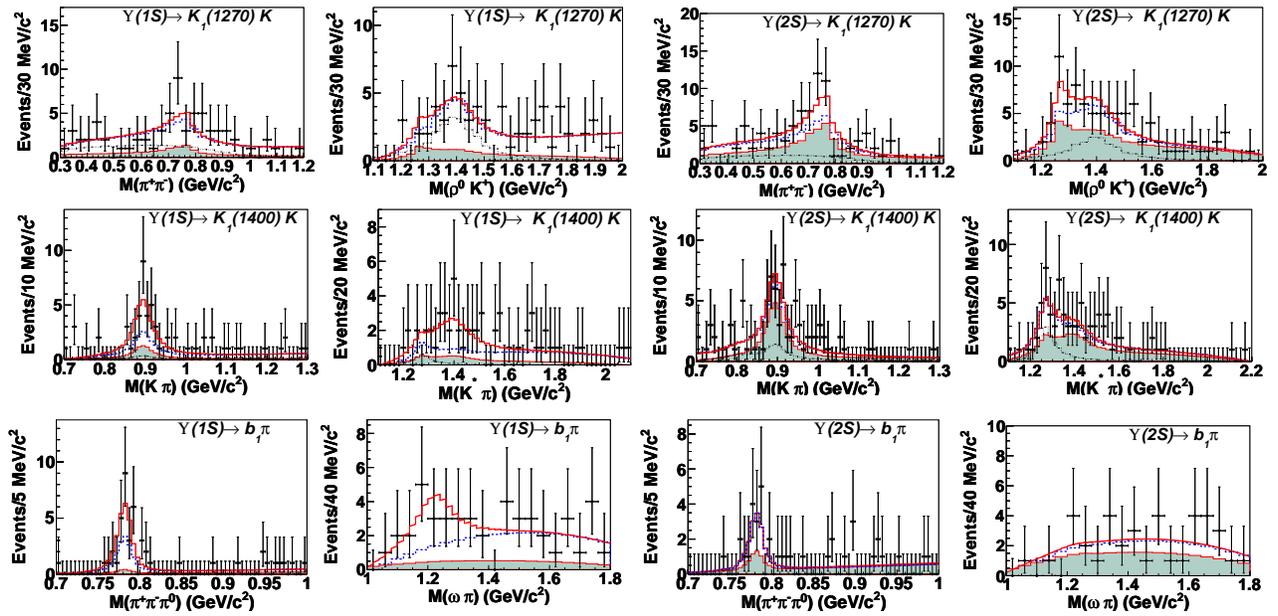

\includegraphics[height=4cm,angle=-90]{fig4a.epsi}\hspace{0.15cm}\vspace{0.05cm}
\includegraphics[height=4cm,angle=-90]{fig4b.epsi}\hspace{0.25cm}
\includegraphics[height=4cm,angle=-90]{fig4c.epsi}\hspace{0.15cm}\vspace{0.05cm}
\includegraphics[height=4cm,angle=-90]{fig4d.epsi}\hspace{0.15cm}\vspace{0.05cm}
\includegraphics[height=4cm,angle=-90]{fig4e.epsi}\hspace{0.15cm}\vspace{0.05cm}
\includegraphics[height=4cm,angle=-90]{fig4f.epsi}\hspace{0.25cm}\vspace{0.05cm}
\includegraphics[height=4cm,angle=-90]{fig4g.epsi}\hspace{0.15cm}\vspace{0.05cm}
\includegraphics[height=4cm,angle=-90]{fig4h.epsi}\hspace{0.15cm}\vspace{0.05cm}
\includegraphics[height=4cm,angle=-90]{fig4i.epsi}\hspace{0.15cm}\vspace{0.05cm}
\includegraphics[height=4cm,angle=-90]{fig4j.epsi}\hspace{0.25cm}\vspace{0.05cm}
\includegraphics[height=4cm,angle=-90]{fig4k.epsi}\hspace{0.25cm}\vspace{0.05cm}
\includegraphics[height=4cm,angle=-90]{fig4l.epsi}\hspace{0.15cm}\vspace{0.05cm}
\caption{\label{ap-fit} The mass projections for
the vector (from axial-vector decay) and axial-vector meson candidates
from 2D fits to the events from $\Upsilon(1S)$ and
$\Upsilon(2S)$ two-body decays ($AP$ modes). The open histograms
show the results of the 2D simultaneous fits, the dotted curves show
the total background estimates, the dot-dashed curves are the
cross-feed backgrounds described in the text, and the shaded
histograms are the normalized continuum contributions.}
\end{figure*}

For the selected events, Fig.~\ref{vt-fit} shows the invariant
mass distributions for the vector and tensor meson candidates, and
Fig.~\ref{ap-fit} shows the invariant mass distributions for the vector (from
axial-vector decay) and axial-vector meson candidates for
$\Upsilon(1S)$ and $\Upsilon(2S)$ two-body decays.
We extend the unbinned simultaneous maximum likelihood fit described above for three-body
decays into a two-dimensional (2D) fit.
We assume the
mass distributions of the $V$ and $T$ particles to be uncorrelated;
thus, the mass distributions in the 2D space can be represented by the
product of two one-dimensional (1D) probability density functions
(pdf). The 2D fitting function is parameterized as
\begin{eqnarray*}
f(M_1,M_2)= N^{\rm sig} s_1(M_1) s_2(M_2)+
N^{\rm bg}_{sb} s_1(M_1) b_2(M_2) \\
 + N^{\rm bg}_{bs} b_1(M_1)
s_2(M_2)  + N^{\rm bg}_{bb}  b_1(M_1)
b_2(M_2),
\end{eqnarray*}
where $s_1 (M_1)$ and $b_1(M_1)$ are the 1D signal and
background pdfs for $V$, respectively, and $s_2 (M_2)$ and $b_2(M_2)$
are the corresponding pdfs for
$T$. Here, the free parameters are the signal yield $N^{\rm sig}$
and  the background yields  $N^{\rm bg}_{sb}$, $N^{\rm bg}_{bs}$, and $N^{\rm bg}_{bb}$.
Similar 2D pdfs are used to fit the vector (in
axial-vector decays) and axial-vector meson candidates for the $AP$
modes.
In these fits, we assume there is no interference between the
signal and other components due to the limited statistics.
 The 1D projections from the 2D fits are
shown in Figs.~\ref{vt-fit} and \ref{ap-fit} with the contribution
from each component indicated. In the fits to the $K_1(1270)^+ K^-$ and
$K_1(1400)^+ K^-$ modes, the cross-feed background components from
$K_1(1400)^+ K^- \to K^{\ast0} K^- \pi^+ $ and $K_1(1270)^+ K^- \to
\rho^0 K^+ K^-$ are also included and shown as dot-dashed
lines. The fit results are shown in Table~\ref{summary2}.

There are several sources of systematic errors for the branching
fraction measurements. The uncertainty in the
tracking efficiency for tracks with angles and momenta
characteristic of signal events is about 0.35\% per track and is
additive. The uncertainty due to particle
identification efficiency is 1\% with an efficiency correction
factor of 0.97 for each pion and 0.8\% with an efficiency correction
factor 0.97 for each kaon, respectively.  The uncertainty in selecting $\pi^0$ candidates is
estimated by comparing control samples of $\eta \to \pi^0 \pi^0
\pi^0$ and $\eta \to \pp \pi^0$ decays in data and amounts to 3.7\%. Errors on the
branching fractions of the intermediate states are taken from the
PDG listings~\cite{PDG}.  According to MC simulation, the trigger
efficiency is greater than 99\%, so that the corresponding uncertainty can be neglected.
We estimate the systematic errors associated with the fitting procedure by changing
the order of the background polynomial and the range of the fit;  the differences in
the fitted results, which are 1.3\%-29\% depending on the final state particles that are
taken as systematic errors. We estimate the systematic errors associated with the resonance
parameters by changing the values of the masses and widths of the resonances by $\pm 1\sigma$;
the differences of 0.6\%-7.3\% in the fitted results are taken as systematic errors.
For the central values of the branching fractions, the average difference between alternative C.M.
energy dependences of the cross section is included as a systematic error due to the uncertainty
of the continuum contribution, which is in the range of 4.2\% to 22\%.
The uncertainty due to
limited MC statistics is at most 0.5\%. Finally, the uncertainties
on the total numbers of $\Upsilon(1S)$ and $\Upsilon(2S)$ events
are 2.2\% and 2.3\%, respectively. Assuming that all of these
systematic error sources are independent, the total systematic
error is 11\%-31\% depending on the final state.

Table~\ref{summary2} shows the  results for the branching
fractions including the upper limits at 90\% C.L. for the channels
with a statistical significance less than 3$\sigma$.
The corresponding ratio of the branching fractions of $\Upsilon(2S)$
and $\Upsilon(1S)$ decay is calculated; in some cases, the
systematic errors cancel. In order to set
conservative upper limits on these branching fractions, the
efficiencies are lowered by a factor of $1-\sigma_{\rm sys}$ in
the calculation, where $\sigma_{\rm sys}$ is the total systematic
error. All the results on the branching fractions, including upper
limits, are below CLEO's preliminary
results~\cite{cleo-ydecay}.

In summary, we have measured $\Upsilon(1S)$ and $\Upsilon(2S)$ hadronic
exclusive decays to three-body final states and two-body
processes.
Signals are observed for the first time in the
$\Upsilon(1S) \to \phi K^+ K^-$, $\omega \pi^+ \pi^-$,
$K^{\ast 0} K^- \pi^+$, $K^{\ast0} K_2^{\ast 0}$ and
$\Upsilon(2S) \to \phi K^+ K^-$, $K^{\ast 0} K^- \pi^+$ decay modes.
Besides $K^{\ast0} K_2^{\ast 0}$,
no other two-body processes are observed in
all investigated final states.
We find that for the
processes $\phi \kk$, $K^{\ast 0} K^- \pi^+ $, and $K^{\ast
0} \bar{K}^{\ast 0}_2(1430)$,  the $Q_{\Upsilon}$ ratios are consistent with
the expected value, while for $\omega \pp$, the measured $Q_{\Upsilon}$  ratio is
$2.6\sigma$ below the pQCD expectation. The results for the other modes
are inconclusive due to low statistical significance. These
results may supply useful guidance for interpreting
violations of the 12\% rule for OZI suppressed decays in the charmonium sector.


We thank the KEKB group for excellent operation of the
accelerator; the KEK cryogenics group for efficient solenoid
operations; and the KEK computer group, the NII, and
PNNL/EMSL for valuable computing and SINET4 network support.
We acknowledge support from MEXT, JSPS and Nagoya's TLPRC (Japan);
ARC and DIISR (Australia); NSFC (China); MSMT (Czechia);
DST (India); INFN (Italy); MEST, NRF, GSDC of KISTI, and WCU (Korea);
MNiSW (Poland); MES and RFAAE (Russia); ARRS (Slovenia);
SNSF (Switzerland); NSC and MOE (Taiwan); and DOE and NSF (USA).



\begin{thebibliography}{**}
\bibitem{charge} Charge-conjugate decays are implicitly assumed
throughout the paper.
\bibitem{foot} The dependence of the cross section on the beam energy
could vary from $1/s$ to $1/s^4$, which is included as a source
of systematic error.
\end{thebibliography}
\end{document}